\documentclass[10pt]{article}
\usepackage[top=0.85in,left=2.75in,footskip=0.75in]{geometry}

\usepackage{amsmath}
\usepackage{amssymb}
\usepackage{txfonts}
\usepackage{multicol}
\usepackage[ justification=justified, singlelinecheck=false]{caption}

\usepackage{color}
\definecolor{orange}{RGB}{255,127,0}
\definecolor{grey}{RGB}{125,125,125}

\usepackage[pdftex]{graphicx}
\usepackage{epstopdf}

\usepackage{cite}

\usepackage[colorlinks]{hyperref}
\hypersetup{colorlinks,linkcolor=blue,citecolor=blue,urlcolor=blue,filecolor=blue,pdftex}

\usepackage[utf8]{inputenc}
\usepackage{textcomp,marvosym}
\usepackage{fixltx2e}
\usepackage{amsmath,amssymb,bm}
\usepackage[right]{lineno}
\usepackage[aboveskip=1pt,labelfont=bf,labelsep=period,justification=raggedright,singlelinecheck=off]{caption}

\makeatletter
\renewcommand{\@biblabel}[1]{\quad#1.}
\makeatother

\date{}

\usepackage{lastpage,fancyhdr,graphicx}
\usepackage{epstopdf}
\fancyheadoffset[L]{2.25in}
\fancyfootoffset[L]{2.25in}

\usepackage{ulem}

\begin{document}
\vspace*{0.35in}

% Please capitalize all terms in the title except conjunctions, prepositions, and articles.
\begin{flushleft}
{\Large
\textbf\newline{Extracranial estimation of neural mass model parameters using the Unscented Kalman Filter}
}
\newline
\\
Lara Escuain-Poole\textsuperscript{1},
Jordi Garcia-Ojalvo\textsuperscript{2,*},
Antonio J. Pons\textsuperscript{1,*}
\\
\bigskip
\textsuperscript{1} Department of Physics, Polytechnic University of Catalonia, Terrassa, Spain
\\
\textsuperscript{2} Department of Health and Experimental Sciences, Pompeu Fabra University, Barcelona, Spain
\bigskip

% Use the asterisk to denote corresponding authorship and provide email address in note below.
* jordi.g.ojalvo@upf.edu, a.pons@upc.edu

\end{flushleft}
\section*{Abstract}

Data assimilation, defined as the fusion of data with preexisting knowledge, is particularly suited to elucidating underlying phenomena from noisy/insufficient observations.
Although this approach has been widely used in diverse fields, only recently have efforts been directed to problems in neuroscience, using mainly intracranial data and thus limiting its applicability to invasive measurements involving electrode implants.
Here we intend to apply data assimilation to non-invasive electroencephalography (EEG) measurements to infer brain states and their characteristics.
For this purpose, we use Kalman filtering to combine synthetic EEG data with a coupled neural-mass model together with Ary's model of the head, which projects intracranial signals onto the scalp.
Our results show that using several extracranial electrodes allows to successfully estimate the state and parameters of the neural masses and their interactions, whereas one single electrode provides only a very partial and insufficient view of the system.
The superiority of using multiple extracranial electrodes over using only one, be it intra- or extracranial, is shown over a wide variety of dynamical behaviours. Our results show potential towards future clinical applications of the method.

\section*{Author Summary}

%\linenumbers

To completely understand brain function, we will need to integrate experimental information into a consistent theoretical framework.
Invasive techniques as EcoG recordings, together with models that describe the brain at the mesoscale, provide valuable information about the brain state and its dynamical evolution when combined with techniques coming from control theory, such as the Kalman filter.
This method, which is specifically designed to deal with systems with noisy or imperfect data, combines experimental data with theoretical models assuming Bayesian inference.
So far, implementations of the Kalman filter have not been suited for non-invasive measures like EEG.
Here we attempt to overcome this situation by introducing a model of the head that allows to transfer the intracranial signals produced by a mesoscopic model to the scalp in the form of EEG recordings.
Our results show the advantages of using multichannel EEG recordings, which are extended in space and allow to discriminate signals produced by the interaction of coupled columns.
The extension of the Kalman method presented here can be expected to expand the applicability of the technique to all situations where EEG recordings are used, including the routine monitoring of illnesses or rehabilitation tasks, brain-computer interface protocols, and transcranial stimulation.

\section*{Introduction}

After several decades studying its morphology and dynamics \cite{Yuste2015}, the basic mechanisms that describe the functioning of the brain are still far from being completely understood.
There are different reasons that explain this arduous route towards understanding this organ.
First, the neurons that form the brain are very diverse morphologically \cite{Braitenberg91a} and dynamically \cite{Izhikevich2008}.
Second, these neurons are connected to each other in extremely large numbers and forming very complex networks \cite{Callaway2016}, whose structural characteristics are still mostly unknown.
And third, brain dynamics are very irregular and complex \cite{Fre87,Faure2001}.
The opposed views of an essentially noisy brain and a deterministic brain exhibiting chaotic activity have been often contrasted.
On the one hand there is multiple evidence, both theoretical and experimental, that justifies a stochastic view of the brain\cite{Shadlen1994,Faisal2008}.
On the other hand, other studies reveal deterministic, or rather reproducible, dynamical behaviour\cite{Schiff1994,VanVreeswijk1996} both at the microscopic scale \cite{Celletti1999} and at the mesoscale recorded by electroencephalograms (EEG) or magnetoencephalograms (MEG) \cite{Stam2005}.
The reality is probably a combination of the two views.
The fact that the brain receives continuous external inputs from the sensory system also makes its dynamical and experimental interpretation more complex because, even though experiments are designed to minimize uncontrolled inputs, they cannot completely rule them out.
Another important limitation for studying the brain is that experimental recordings (such as EEG or fRMI) are almost always indirect reflections of the underlying neural activity \cite{Buzsaki2012}.

A way of facing the complexities described above is by systematically comparing the experimental observations of brain activity with mathematical models based on specific hypotheses, which can thereby be validated or disproven.
Modelling cerebral activity has been attempted both with top-down and bottom-up approaches \cite{Wright1996,Rabinovich2006,Eliasmith2012,Deco2014}.
Many of these theoretical models are simplifications that capture the basic ingredients of brain dynamics, while others are detailed accounts of the dynamics of neurons that necessarily forgo the description of the whole brain.
In that context, a more feasible scale of study is the mesoscopic scale \cite{David2003b,Grimbert2006a,Cona2011a,Coombes2010,Babajani2006a,Babiloni2005,Bojak2010}.
Many of the modern experimental techniques record information coming from populations of neurons organized in so-called cortical columns.
Neural mass models describe mathematically the activity of these populations using reasonably simple equations \cite{Jan95}. These models can describe both the intrinsic oscillatory behaviour recorded at the mesoscale or event-related responses \cite{David2005,Spiegler2011} with morphologically plausible assumptions for their construction.

In all modeling strategies, however, identifying realistic values for the parameters of the model is a challenging task.
One way to address this problem is by integrating experimental information into the models using Bayesian inference\cite{Wright2001,Schiff2008,Kiebel2008,Shine2015,Ma2016}.
This strategy has started to be pursued by using Kalman filtering to integrate experimental data at both the microscopic scale of neuronal networks \cite{Hamilton2014} and the mesoscopic scale of neural mass models \cite{Freestone2013b}.
This data assimilation approach is based on the fact that neuronal activity is highly noisy, and allows to estimate both the state and the parameters of the theoretical model using the experimental data available.
The method has been used to estimate, for example, the effective connectivity that characterizes epileptic seizures on a patient-specific basis (see \cite{Freestone2014} and references therein).
Kalman filtering has also been used to analyze the suppression of epileptic seizures in coupled neural mass models\cite{Cao2015}, and the induction of the anaesthetised state by drugs \cite{Bojak2005,Kuhlmann2016}.
But these studies use mainly invasive intracranial signals, and it would be desirable to extend them to non-invasive extracraneal measurements such as EEG.
Intracranial signals can be translated into EEG signals in a forward manner\cite{Zhang1995,Mosher1999}, and in the opposite direction, solving the inverse problem allows to infer intracranial signals from EEG recordings \cite{Haufe2011,Verhellen2007,Gotman2003}.
In this paper we are going to adopt this approach to extend the Kalman filtering technique, by including a model of the head that transfers intracranial signals onto the scalp.

\section*{Methods}\label{methods}

To obtain a reliable estimation of the state and the dynamics of the brain, we require a biologically inspired mathematical model of its dynamics, experimental data (as non-invasive as possible), and the means of fusing both sources of information together.
In this paper, we use Jansen and Rit's model~\cite{Jan93,Jan95} to represent the dynamical evolution of the cortical structures.
For our sets of experimental data we model EEG \textit{in silico} using, again, Jansen and Rit's model together with a model of the head. Finally, we use the Unscented Kalman Filter as our data assimilation algorithm to estimate jointly the state and the parameters of the model \cite{Mer01,Jul04,Kal60}.

\subsection*{Mesoscopic neural mass model}

Jansen and Rit's model~\cite{Jan93,Jan95} describes the mesoscopic activity of a population of neurons \cite{Lop74,Fau09}, providing a good compromise between physiological realism and computational simplicity. This model simplifies the neuronal diversity of a cortical column in three interacting populations: pyramidal neurons, excitatory interneurons, and inhibitory interneurons. The larger pyramidal population excites both groups of interneurons, which in turn feed back into the pyramidal cells. In our approximation, the pyramidal population is also driven by excitatory noise from distant areas of the brain and by neighbouring columns.

The dynamics of each population rely on two different transformations. The first converts the average density of incoming action potentials into an average post-synaptic membrane potential (excitatory or inhibitory). It takes the form of a second-order differential equation for excitatory inputs,
\begin{equation}
\ddot{x}_e(t) + 2a\dot{x}_e(t) + a^2x_e(t) = Aa\ u_e(t)\label{eq:2ndorder1},
\end{equation}
and for inhibitory inputs,
\begin{equation}
\ddot{x}_i(t) + 2b\dot{x}_i(t) + b^2x_i(t) = Bb\ u_i(t)\label{eq:2ndorder3},
\end{equation}
where $u_{e,i}(t)$ and $x_{e,i}(t)$ are the input and output of the transformations, respectively, $A$ and $B$ are the amplitudes of the excitatory and inhibitory post-synaptic potentials, and $a$ and $b$ are the lumped representations of the sums of the reciprocal of the time constant of the passive membrane, and all other spatially distributed delays in the dendritic network.

The second transformation converts the net average membrane potential of the population, $v$, into an average firing rate, and is described by the following sigmoid function:
\begin{equation}
{\rm Sigm}(v) = \frac{2 e_0}{1+e^{r(v_0-v)}}
\end{equation}
where $e_0$ is the maximum firing rate of the population, $r$ controls the slope of the sigmoid, and $v_0$ is the post-synaptic potential for which a 50\% firing rate is obtained.

We model the brain as a system of $N_d$ coupled cortical columns (dipole sources) with the addition of noise. The following equations define our model for each cortical column $i$:
\begin{align}
\ddot{x}_0^i(t) + 2a\dot{x}_0^i(t) + a^2x_0^i(t) = &Aa\ {\rm Sigm}[x_1^i(t) - x_2^i(t)], \label{Eq:Jansen_01}\\
\ddot{x}_1^i(t) + 2a\dot{x}_1^i(t) + a^2x_1^i(t) = &Aa\ \left(p(t) + k\sum_{j=1}^{N_d} K^{ij}\ {\rm Sigm}(x_1^j(t-\tau^{ij}) - x_2^j(t-\tau^{ij})) \right. \nonumber\\
&\quad + \left.C_2\ {\rm Sigm}[C_1 x_0^i(t)]\vphantom{k\sum_{j=1}^{N} K_{ij}\ {\rm Sigm}(x_1^j(t-\tau)}\right), \label{Eq:Jansen_02}\\
\ddot{x}_2^i(t) + 2b\dot{x}_2^i(t) + b^2x_2^i(t) = &Bb \left(C_4\ {\rm Sigm}[C_3 x_0^i(t)]\right),\label{Eq:Jansen_03}
\end{align}
where $C_1$ to $C_4$ are connectivity constants that govern the interactions between populations, $p(t)$ is a noisy external input, and the summation term includes the delayed input from other coupled cortical columns. $k$ modulates the strength of the coupling, $K$ is the adjacency matrix, and $\tau^{ij}$ is the delay with which column $i$ receives the signal of column $j$. Table~\ref{table:parameters} provides the descriptions and values of these parameters. The electrical activity detected by the electrodes on the scalp is originated by the weighted sum of the averaged membrane potential of the pyramidal cells of all the cortical columns, $x^i(t) = x_1^i(t)-x_2^i(t)$ \cite{Lop10}.

\begin{table}[!ht]
\begin{center}
%\begin{adjustwidth}{-2.25in}{0in} % Comment out/remove adjustwidth environment if table fits in text column.
\caption{
{\bf Description and default values of the parameters for the system of neural masses.} See the~\nameref{results} section for details of the configuration of each numerical experiment. Here, PC refers to pyramidal cells, EI to excitatory interneurons, II to inhibitory interneurons, EPSP to excitatory post-synaptic potential, and IPSP to inhibitory post-synaptic potential.}
\begin{tabular}{|c|l|l|}
\hline
\bf{Param.} & \bf{Description} & \bf{Value} \\ \hline
$A$ & EPSP amplitude & $3.25~mV$ \\ \hline
$B$ & IPSP amplitude & $22.00~mV$ \\ \hline
$a$ & Rate constant for the excitatory population* & $100~s^{-1}$ \\ \hline
$b$ & Rate constant for the inhibitory population* & $50~s^{-1}$ \\ \hline
$C_1$ & Strength of synaptic connections from PC to EI & $135$ \\ \hline
$C_2$ & Strength of synaptic connections from II to PC & $108$ \\ \hline
$C_3$ & Strength of synaptic connections from PC to II & $33.75$ \\ \hline
$C_4$ & Strength of synaptic connections from EI to PP & $33.75$ \\ \hline
$e_0$ & Maximum firing rate of the population & $2.5~s^{-1}$ \\ \hline
$v_0$ & Mean threshold of the population & $6~mV$ \\ \hline
$r$ & Steepness of the sigmoidal transformation & $0.56~mV^{-1}$ \\ \hline
$k$ & Coupling constant & $10$ \\ \hline
$K$ & Adjacency matrix & $K_{ij} = 1, i \neq j$ \\
 &  & $K_{i,j} = 0, i = j$ \\ \hline
$\tau$ & Delay & According  \\
 &  &  to distance~\cite{Pons2010} \\ \hline
$p$ & External input & $200~s^{-1}$ \\ \hline
\end{tabular}
\begin{flushleft} *Lumped representation of the sum of the reciprocal of the time constant of passive membrane and all other spatially distributed delays.
\end{flushleft}
\label{table:parameters}
\end{center}
%\end{adjustwidth}
\end{table}

\subsection*{Extracranial data generation}\label{head}

\begin{figure}[!htb]
\begin{center}
\includegraphics[width=\textwidth]{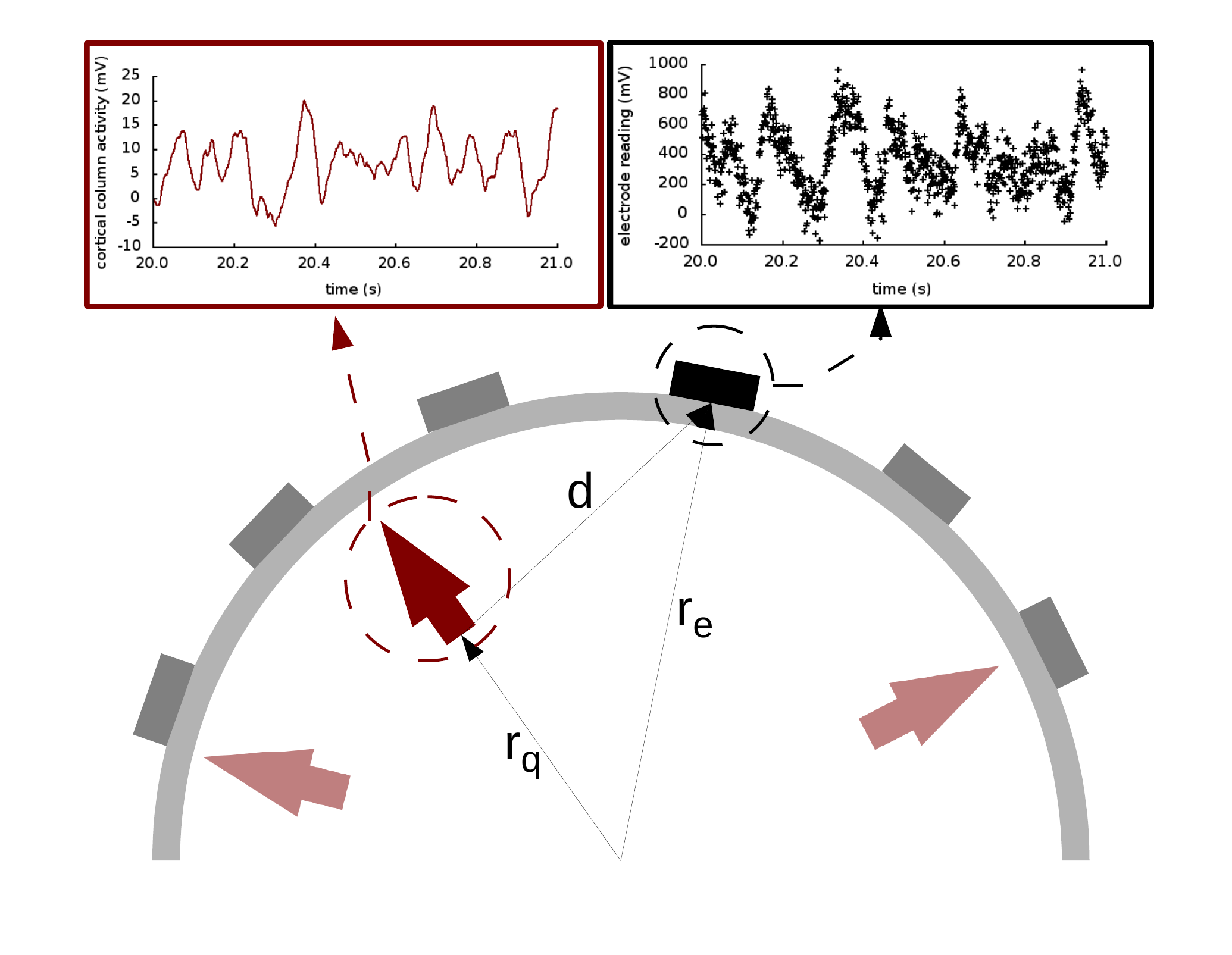}
\caption{{\bf Extracranial data generation and illustration of Ary's model of the head.}
The light and dark red arrows indicate dipole sources, and the electrodes are shown as grey and black rectangles. The elements in the cartoon illustrate how all the signals produced by the cortical columns (represented with the solid red line in the top left panel) are transformed into an electrode reading (shown in black dots in the top right panel) through the lead field matrix. In this drawing, as in Equations~(\ref{eq:lead_field_matrix_first}-\ref{eq:lead_field_matrix_last}), $\bf{r_q}$ is the distance from the origin to the cortical column under consideration; $\bf{r_e}$ is the distance from the origin to the electrode; and $\bf{d}$ is the distance from the cortical column to the electrode. The placement of the arrows here is for illustration purposes only; in our study, the cortical columns are placed on the surface of the brain, close to the skull.}
\label{idea}
\end{center}
\end{figure}

The main contribution of this paper is the use of multichannel extracranial data to obtain information about the neuronal populations inside the brain using data assimilation.
To accomplish this, we use synthetic EEG data generated \textit{in silico} using Jansen and Rit's model and Ary's head model.
To that end, we transform the output $\bm{x}(t)$ of the neural masses to EEG signals $\bm{z}(t)$ in the electrodes (see Fig.~\ref{idea}).
This transformation is mediated by a lead field matrix \cite{Mosher1999}, which builds on the basic idea of calculating the electric potential caused by a dipole source \cite{Buzsaki2012} on a three-layer isotropic hemisphere \cite{Zhang1995,Ary81}.
The lead field matrix also contains information about the geometry of the problem (e.g., locations of cortical columns and electrodes) and about the electrophysiology of the head (e.g., conductivities of the different tissues). The following equations show the potential $V^{e,i}$ on an electrode $e$, located at $\bm{r_e}^e$~\cite{Jurcak2007}
, caused by the dipole $\bm{q}^i(t)=x^i(t) \bm{\hat{q}}^i$ generated by the cortical column $i$, located at $\bm{r_q}^i$ and oriented as $\bm{\hat{q}}^i$. In these equations, $e = 1,\dotsc, N_e$, where $N_e$ is the total number of electrodes, and $i = 1,\dotsc, N_d$, where $N_d$ is the total number of dipoles:
\begin{align}
&V^{e,i}(\bm{r}_{\bm{e}}^e; \bm{r}_{\bm{q}}^i, \bm{q}^i) \approxeq v^{1}(\bm{r}_{\bm{e}}^e; \mu_1\bm{r}_{\bm{q}}^i, \rho_1\bm{q}^i) + v^{2}(\bm{r}_{\bm{e}}^e; \mu_2\bm{r}_{\bm{q}}^i, \rho_2\bm{q}^i) + v^{3}(\bm{r}_{\bm{e}}^e; \mu_3\bm{r}_{\bm{q}}^i, \rho_3\bm{q}^i),\label{eq:lead_field_matrix_first}
\end{align}
where vectors are typeset in bold and
\begin{align}
&v^{1}(\bm{r}_{\bm{e}}^e; \bm{r}_{\bm{q}}^i, \bm{q}^i) = \left((c_1^{e,i,1} - c_2^{e,i,1}(\bm{r}_{\bm{e}}^e\cdot\bm{r}_{\bm{q}}^i) )\bm{r}_{\bm{q}}^i + c_2^{e,i,1}(r_q^i)^2\bm{r}_{\bm{e}}^e\right)\cdot \bm{q}^i, \\
&v^{2}(\bm{r}_{\bm{e}}^e; \bm{r}_{\bm{q}}^i, \bm{q}^i) = \left((c_1^{e,i,2} - c_2^{e,i,2}(\bm{r}_{\bm{e}}^e\cdot\bm{r}_{\bm{q}}^i) )\bm{r}_{\bm{q}}^i + c_2^{e,i,2}(r_q^i)^2\bm{r}_{\bm{e}}^e\right)\cdot \bm{q}^i, \\
&v^{3}(\bm{r}_{\bm{e}}^e; \bm{r}_{\bm{q}}^i, \bm{q}^i) = \left((c_1^{e,i,3} - c_2^{e,i,3}(\bm{r}_{\bm{e}}^e\cdot\bm{r}_{\bm{q}}^i) )\bm{r}_{\bm{q}}^i + c_2^{e,i,3}(r_q^i)^2\bm{r}_{\bm{e}}^e\right)\cdot \bm{q}^i.
\end{align}
In these expressions,
\begin{align}
\begin{split}
&c_1^{e,i,s} = \frac{1}{4\pi\sigma^s (r^i_q)^2}\left(2 \frac{\bm{d}^{e,i}\cdot\bm{r_q}^i}{(d^{e,i})^3} + \frac{1}{d^{e,i}} - \frac{1}{r_e^e}\right), \\
&c_2^{e,i,s} = \frac{1}{4\pi\sigma^s (r^i_q)^2}\left(\frac{2}{(d^{e,i})^3} + \frac{d^{e,i}+r_e^e}{r_e\Gamma(\bm{r_e}^e, \bm{r_q}^i)} \right), \\
&\Gamma(\bm{r_e}^e, \bm{r_q}^i) = d^{e,i}\left(r_e^e d^{e,i} + (r_e^e)^2 - (\bm{r_q}^i\cdot\bm{r_e}^e)\right).\label{eq:lead_field_matrix_last}
\end{split}
\end{align}
$\sigma^s$ is the tangential conductivity of each surface~\cite{Ary81} and $\rho_s$ and $\mu_s$ are the Berg parameters relative to it~\cite{Ber94} (see Table~\ref{berg_parameters}). $\bm{d}^{e,i}$ is the distance between the dipole $i$ and the electrode $e$ under consideration.

\begin{table}[!ht]
\begin{center}
\caption{
{\bf Values of the Berg parameters for the three surfaces~\cite{Ary81,Ber94}.}}
\begin{tabular}{|l|l|l|l|}
\hline
\textbf{parameter} & \textbf{surface 1} & \textbf{surface 2} & \textbf{surface 3} \\ \hline
Tangential conductivity $\sigma^s$ & 1.0 & 0.0125 & 1.0 \\ \hline
Berg parameter $\rho_s$ & 0.9901 & 0.7687 & 0.4421 \\ \hline
Berg parameter $\mu_s$ & 0.0659 & 0.2389 & 0.3561 \\ \hline
\end{tabular}
\label{berg_parameters}
%\end{adjustwidth}
\end{center}
\end{table}

\begin{table}[!htb]
\begin{center}
\caption{
{\bf Cartesian coordinates of the dipoles used throughout the study.}}
\begin{tabular}{|l|l|l|l|}
\hline
\textbf{} & \textbf{x} & \textbf{y} & \textbf{z} \\ \hline
dipole 1 & 0.1688 & 0.2242 & 0.2597 \\ \hline
dipole 2 & 0.3766 & -0.8520 & 0.2597 \\ \hline
dipole 3 & 0.6622 & -0.2242 & -0.1948 \\ \hline
\end{tabular}
\label{dipole_coordinates}
%\end{adjustwidth}
\end{center}
\end{table}

\subsection*{The Unscented Kalman Filter for data assimilation}

The Unscented Kalman Filter (UKF) is our algorithm of choice to bring together the dynamical state of the model and the experimental data. It is a standard tool in the field of systems and control engineering, and has been shown to be both computationally efficient and robust even when dealing with stochastic nonlinear systems~\cite{Mer00}. In order to simultaneously estimate the state and parameters of the model described by Eqs.~(\ref{Eq:Jansen_01})-(\ref{Eq:Jansen_03}), we regard it as a discrete-time state-space dynamical system of the following form:
\begin{align}
\bm{x}_{k+1} &= \bm{F}\left(\bm{x}_k\right) + \bm{v}_k \label{eq:state} \\
%\bm{\theta}_{k+1} &= \bm{\theta}_k + \bm{g}_k \\
\bm{z}_k &= \bm{H}\left(\bm{x}_k\right) + \bm{w}_k \label{eq:measurement}
\end{align}
where $\bm{x} = (x_0^1, x_1^1, x_2^1, x_0^2,\dotsc,x_2^{N^d}, \theta^1,\dotsc,\theta^{N_p}) \in \mathbb{R}^{n_x}$ is the state vector (related to the variables and parameters of the model), with $\theta^p$ being the parameters to estimate, which obey the equations $\dot{\theta}^p = 0$. $\bm{z} \in \mathbb{R}^{n_z}$ is the measurement vector (our \textit{in silico} EEG readings). $\bm{v}$ and $\bm{w}$ are uncertainty terms that account for process noise and measurement noise, respectively, with Gaussian distributions $p(\bm{v}) \sim N(0, \bm{Q})$ and $p(\bm{w}) \sim N(0, \bm{R})$, respectively. $\bm{F}$ is obtained with a numerical implementation of Eqs.~(\ref{Eq:Jansen_01})-(\ref{Eq:Jansen_03}), as described below. $\bm{H}$ relates the state to measurement space. Interestingly, this basic part of the Kalman filter is in our case implemented by the skull, the effect of which is represented by the lead field matrix, based on Ary's head model and introduced above.

The UKF is a recursive predictor-corrector-type algorithm that aims to minimise the mean square error of the estimated states and parameters over time. For each time step it calculates a prediction of the state and parameters of the system, which is corrected when the information from a measurement is incorporated. The amount of confidence given to the model and measurement is quantified by the Kalman gain $\bm{K}$, which is calculated at each time step based on prediction covariances as well as model and measurement error covariances ($\bm{Q}$ and $\bm{R}$, respectively). For more details on the implementation of the filter, the reader is referred to~\nameref{s1_appendix} and to Refs.~\cite{Kal60,Mer01,Jul04,Solonen2014}.

\subsection*{Generation of \textit{in silico} datasets}

\begin{figure}[!htb]
\begin{center}
\includegraphics[width=4cm]{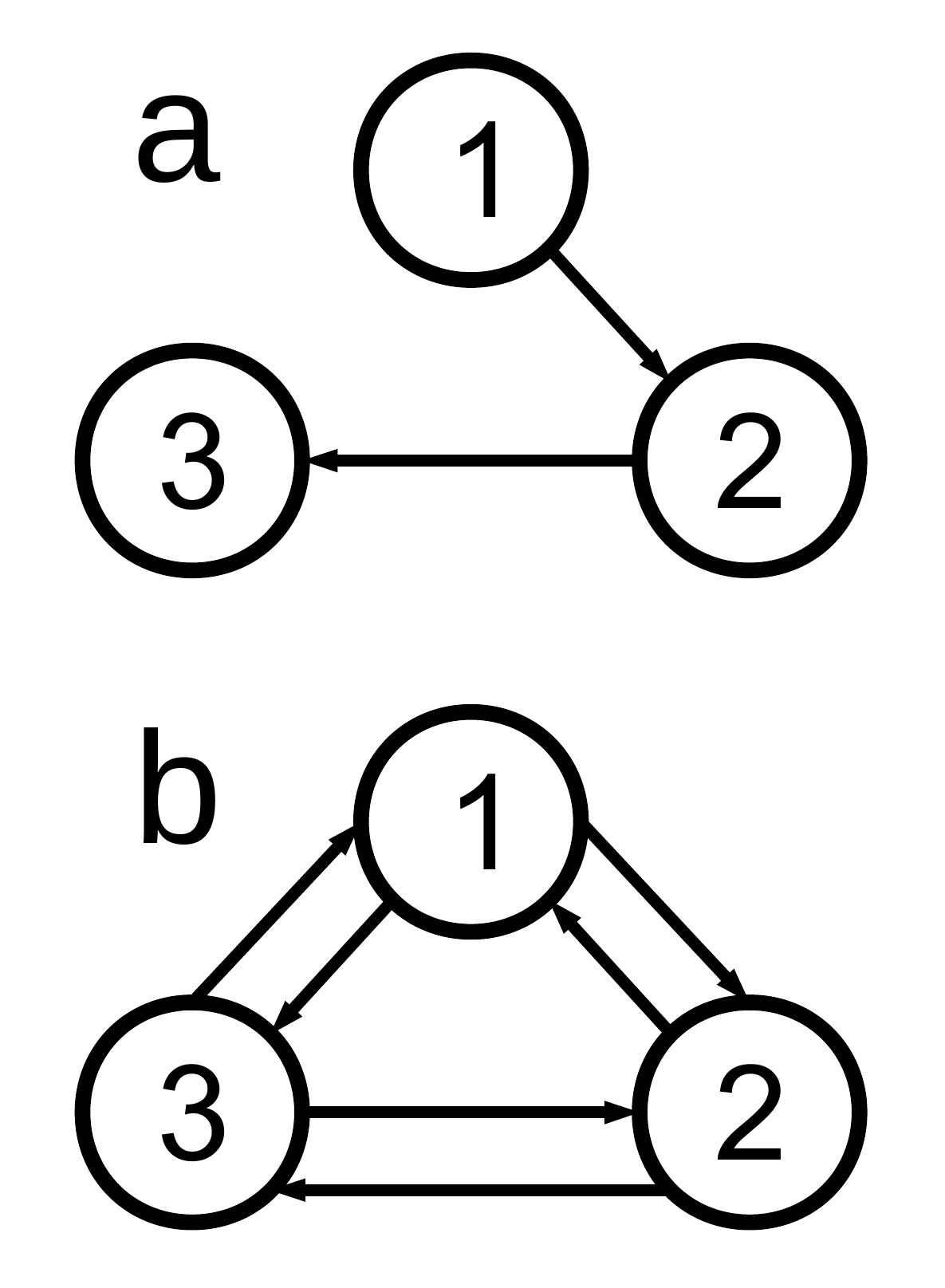}
\caption{{\bf The two cortical column motifs used in this paper.}
Unidirectionally coupled cortical columns have no backflow, and bidirectionally coupled columns are coupled all-to-all. See Table~\ref{dipole_coordinates}.}
\label{configurations}
\end{center}
\end{figure}

For this paper three different \textit{in silico} datasets were generated. We consider both simulated electrocorticography (ECoG, intracortical) and electroencephalography (EEG, extracranial) readings (using Ary's model in the latter case). All datasets used the same locations~\cite{Ric04} for cortical columns and electrodes, as shown in Figs.~\ref{histograms_1} to~\ref{histograms_3}. The strength of the coupling was set at a medium value so that the cortical columns have an effect on one another without fully synchronizing behaviours, and the configurations of the couplings are as shown in Fig.~\ref{configurations}. Table~\ref{table:parameters} shows representative values for the parameters used in all analyses unless otherwise specified. In this paper we focus on estimating the amplitudes $A$ of the EPSPs of the different cortical columns, and therefore we choose values for these amplitudes that produce signals that reflect various dynamic regimes that we wish to explore. The numerical solver used to generate the {\it in silico} time
series was the Heun algorithm~\cite{Toral2014} with a time step of $\Delta t = 1$~ms; the length of the data is of $100$~s in all cases.

To set the matrices $\bm{Q}$ and $\bm{R}$---which reflect the quality of measurement and model, and which crucially affect the output of the filter---, we used our knowledge of the characteristics of the data to fix an initial guess~\cite{Liu2013}, then adjusted it to meet performance criteria.

For each of the experiments we conducted 50 realizations of each estimation with different initial conditions; therefore, all the figures show averages of the 50 estimations, unless otherwise specified. The initial conditions for state and parameter estimations were randomly generated; the parameters, however, were constrained to deviate no more than 90\% of their actual value as an initial assumption.

% Results and Discussion can be combined.
\section*{Results}\label{results}

In order to compare the performance of the extra- and intracranial approaches to Kalman filtering, we have analyzed three different cortical column configurations, each using one of the two motifs shown in Fig.~\ref{configurations}.
Where relevant, two different types of estimations have been used: intracranial and extracranial.
Intracranial estimation uses experimental data that would have hypothetically been obtained from an electrocorticography, that is, using a single intracortical electrode, and is estimated with the data provided by a single location ---in other words, the direct output of Jansen and Rit's model.
Extracranial estimation, on the other hand, employs experimental data originated from EEG recordings, using several electrodes placed on the skull, and is implemented here with the projection on the head of the model output. We now discuss the three different dipole configurations that we have considered.

\subsection*{Three unidirectionally coupled cortical columns}

\begin{figure}[!htb]
\begin{center}
\includegraphics[width=12cm]{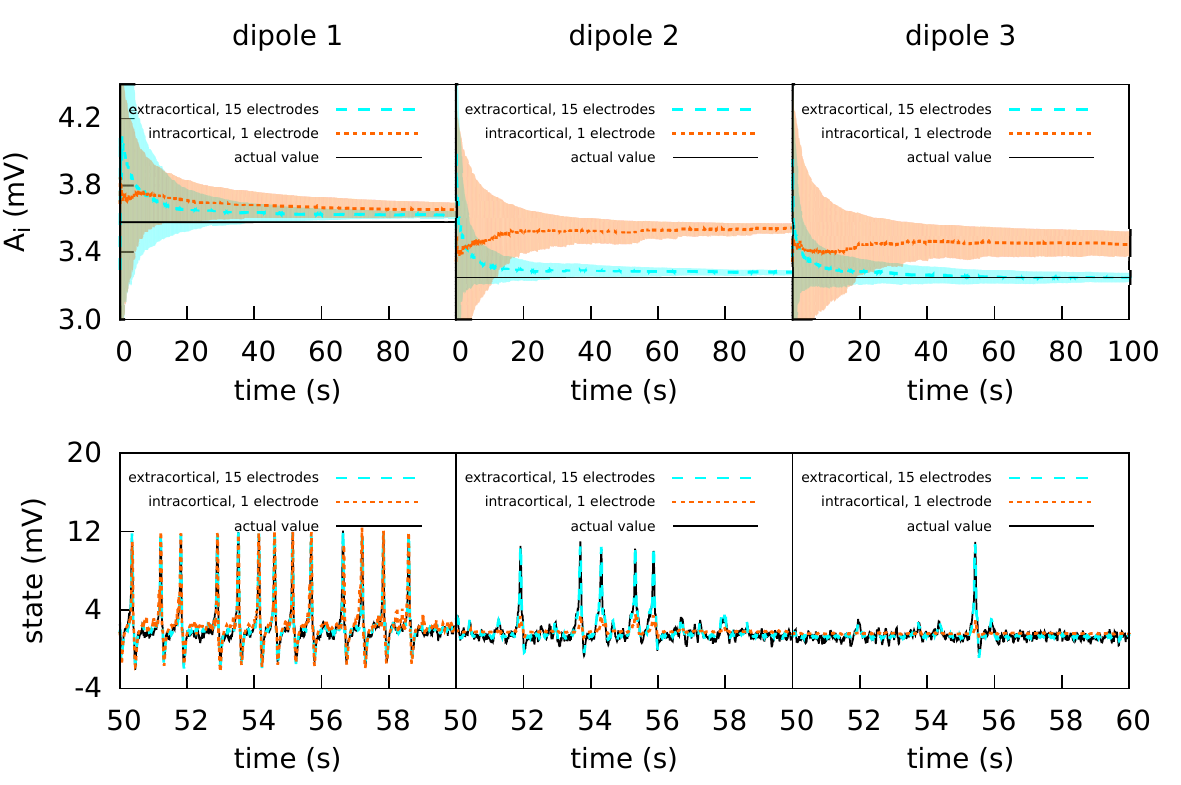}
\caption{{\bf Intracranial and extracranial fittings with propagated excitation along unidirectionally coupled cortical columns.}
The upper panels show the estimation of parameter $A$, and the lower panels show the estimations of the observed states. The solid lines show the averages of the 50 realisations of the estimation, and the shadowed areas indicate the standard deviation. The actual values of the $A$ parameters are $A_1=3.58~mV$, $A_2=3.25~mV$, and $A_3=3.25~mV$, the other parameters being set to standard values (Table~\ref{table:parameters}). All three cortical columns received an external input, $p$, in the shape of Gaussian white noise with mean $90~s^{-1}$ and standard deviation $20~s^{-1}$. The coupling constant was set to $k=10$. The measurements were corrupted with noise of mean $0$ and standard deviation $100~mV$ for extracranial measurements and $5~mV$ for intracortical measurements. Except for cortical column 1, with intracortical data the filter converges to a much higher value than the target, whereas with extracranial data the filter converges to a value which is accurate. In the lower panels it is shown that
extracranial estimations of the state are also accurate, whereas intracortical estimations fail to reproduce the spikes correctly.}
\label{exp_08}
\end{center}
\end{figure}

The first study was performed with the cortical columns coupled unidirectionally (panel (a) of Fig.~\ref{configurations}), as described in~\cite{Liu2013}. The parameters were set to standard values~\cite{Jan95} for the three cortical columns (see Table~\ref{table:parameters}), except for the first column, in which $A_1$ was set to 3.58~mV to make it hyperexcitable. The coupling constant was set to a medium value, large enough for the cortical columns to have a visible effect on each other but not so large that they will fully synchronize and lock their dynamics. In this case, information flows unidirectionally because of the way the cortical columns are coupled~\cite{Liu2013}. As can be seen in the lower panels of Fig.~\ref{exp_08}, the first cortical column has a random spiking activity, due to the increased value of $A$ and the presence of noise~\cite{Gri06}.
Due to the architecture of the coupling, cortical column 1 causes cortical columns 2 and 3 to spike also, when otherwise they would have simply fluctuated around their resting level.

The upper panels of Fig.~\ref{exp_08} show the intracortical and extracranial estimations of $A$ for the three cortical columns. The estimation for $A_1$ of the first column converges to its correct value, with both the intra- and extracortical approaches.
This was to be expected, since the first cortical column receives no inputs from other elements of the system. In contrast, the intracortical estimations for cortical columns 2 and 3 converge to values significantly higher than their actual value of $3.25~mV$.
We conjecture that this is caused by the spiking of these two cortical columns, which as mentioned above is due to the influence of cortical column 1.
Multi-channel extracranial information, in contrast, allows to see the complete picture of the coupled cortical columns and treat them as a single composed system, contrary to the partial picture obtained from the information provided by the single intracranial recordings.
Therefore, estimation is better when using extracranial information with several electrodes, as shown in the upper panels of the figure.
The lower panels of Fig.~\ref{exp_08} show the estimation of the state. The UKF shows great efficacy when the estimation is extracranial, but performs poorly in the case of intracortical estimation (with the exception of cortical column 1, because it has no input from other cortical columns).

\subsection*{Three bidirectionally coupled cortical columns: coarse parameter estimation}
The second experiment aims to explore the filter's possibilities in more extreme situations. The three cortical columns are located as in the previous section, but coupled bidirectionally (panel (b) of Fig.~\ref{configurations}). Additionally, the maximum amplitudes of the excitatory PSPs are set to $A_1 = 4.25~mV$, $A_2 = 10.00~mV$, and $A_3 = 3.25~mV$. These values were chosen to force the three cortical columns to be in very different dynamical regimes: cortical column 1 operates in a spiking regime; cortical column 2 oscillates with alpha frequency but with an amplitude similar to that of the spikes; and cortical column 3 oscillates in a more standard regime, as described in \cite{Jan95}.

\begin{figure}[!htb]
\begin{center}
\includegraphics[width=12cm]{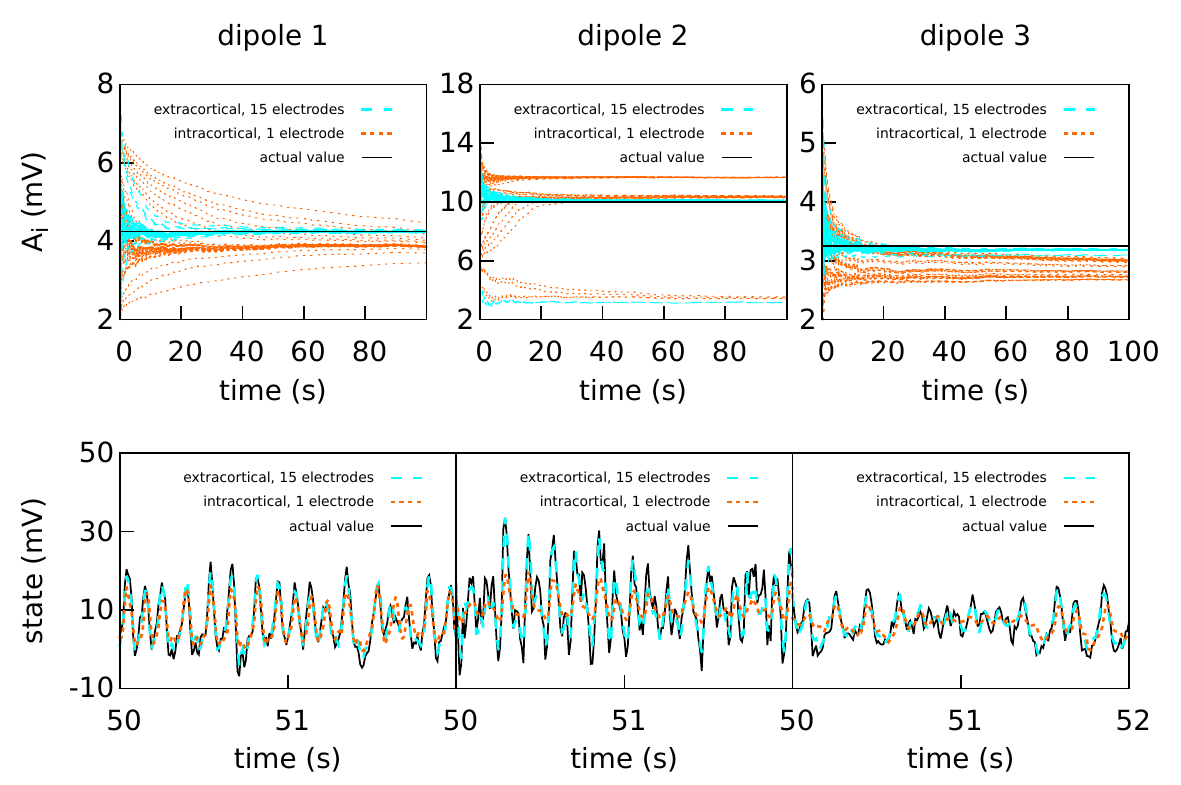}
\caption{{\bf Intracranial and extracranial fittings for coarse parameter estimation in the case of bidirectional coupling.}
As in the previous figure, the upper panels show the estimation of $A$ for each cortical column and the lower panels show the estimations of the observed states. The results are shown here without averaging. The actual values of the amplitudes of the EPSPs are $A_1 = 4.25~mV$, $A_2 = 10.00~mV$, and $A_3 = 3.25~mV$; the rest of the parameters were set to standard values (Table~\ref{table:parameters}). The external input $p$ for each of the three cortical columns is of mean $200~s^{-1}$ and standard deviation $100~s^{-1}$. The coupling constant was set to $k=5$. The intracortical measurements were corrupted with noise of mean $0$ and standard deviation $5~mV$, while the noise in the extracranial measurements is of standard deviation $100~mV$. Extracranial estimations of the parameters are both faster and more accurate than intracortical estimations; this applies also to the state, whose dynamics are more faithfully reproduced using multi-electrode extracranial estimation (as shown in the lower panels).}
\label{exp_07}
\end{center}
\end{figure}

\paragraph*{Moderate intracortical measurement noise.}
Figure~\ref{exp_07} shows again the performance obtained using the experimental data from a set of extracraneal electrodes compared to using individual intracortical electrodes for each cortical column. In this case we show the 50 realisations of each filtering, without showing the average. The extracranial data for this experiment were corrupted with a measurement noise of zero mean and standard deviation $100~s^{-1}$; the intracortical data were corrupted with a measurement noise of standard deviation $5~s^{-1}$ in order to maintain similar levels of signal-to-noise ratio.

As shown in Fig.~\ref{exp_07}, the intracortical parameter estimations do not approximate very well the target value. In particular, the estimations of $A$ for cortical column 2 converge to three different values depending on the initial conditions. The state estimation follow the actual state of the system closely only for cortical column 1.
The situation is very different when with extracranial electrodes, where all 50 realisations of the estimations converge with much more precision to the correct values for both state and parameters (with the exception of $A_2$, which still tends to lower values in a very small quantity of the realisations). Again, extracranial performance is better, in general, to intracortical.

\begin{figure}[!htb]
\begin{center}
\includegraphics[width=12cm]{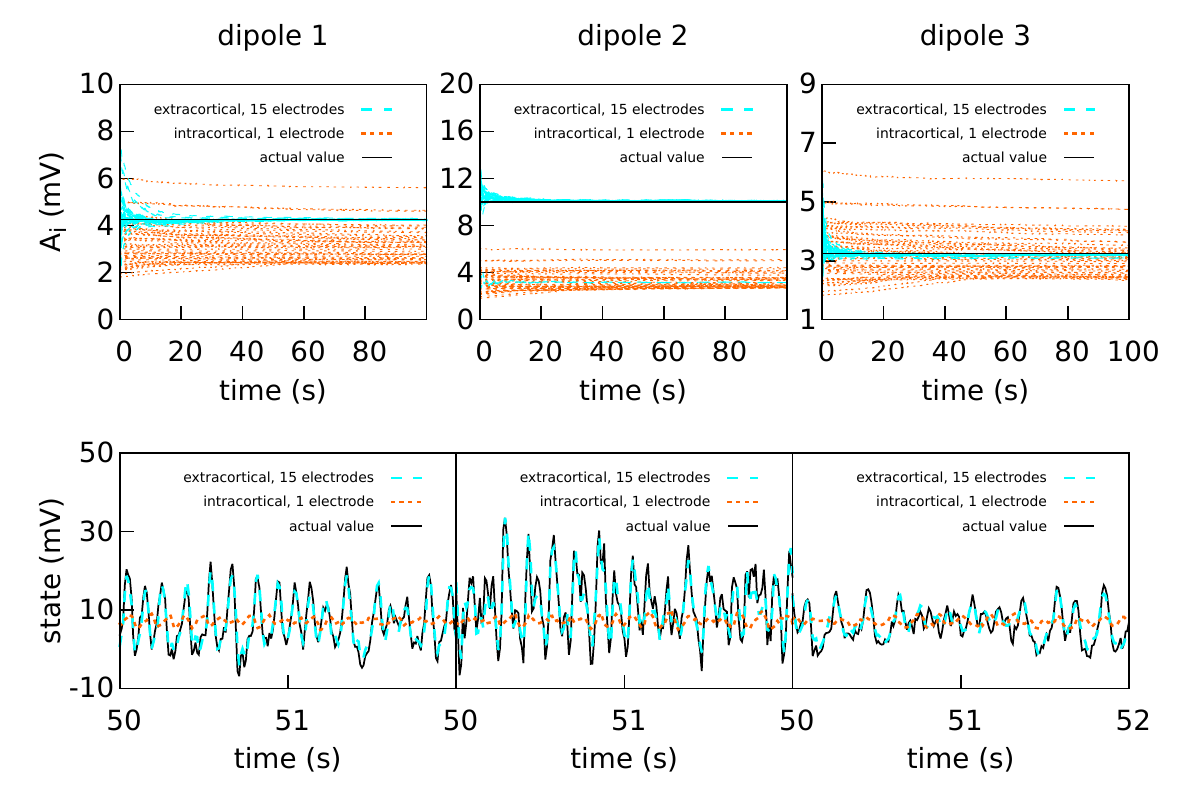}
\caption{{\bf Intracranial and extracranial fittings for coarse parameter estimation, with a higher amount of intracortical measurement noise.}
The upper panels show the estimation of the EPSPs for each cortical column and the lower panels show the estimations of the observed states. The results are shown here without averaging. The actual values of the amplitudes of the EPSPs are $A_1 = 4.25~mV$, $A_2 = 10.00~mV$, and $A_3 = 3.25~mV$; the rest of the parameters were set to standard values (Table~\ref{table:parameters}). The external input $p$ for each of the three cortical columns is of mean $200~s^{-1}$ and standard deviation $100~s^{-1}$. The coupling constant was set to $k=5$. The intracortical measurements were corrupted with noise of mean $0$ and standard deviation $100~mV$---about an order of magnitude higher than the noise in the previous graph---, while the noise in the extracranial measurements is of standard deviation $100~mV$. Extracranial estimations of the parameters are also faster and more accurate than intracortical estimations, more markedly so in this case; as to the state, in this more extreme case, the intracortical estimation
does not mimic the evolution of the system in any way.}
\label{exp_07_noisier}
\end{center}
\end{figure}

\paragraph*{High intracortical measurement noise.}
The difference between intracranial and extracranial estimation is even larger for higher measurement noise (Fig.~\ref{exp_07_noisier}). In this case, the amount of noise in the intracortical data was set to the same value as the noise in the extracranial data. The value of $\bm{R}$ was tuned to reflect the increase in measurement noise, but the intracortical estimations failed to obtain the correct values for the parameters and reproduce the state.

\paragraph*{Using one single extracranial electrode.}
Using the same dataset, we aimed to investigate the outcome of using each extracranial electrode individually~\cite{Fre14}, as opposed to using the complete subset as until now. Therefore we used each electrode separately to estimate the state and parameters of the complete system, with 50 realisations of the estimation for each electrode. By doing so, we show that the quality of the estimations is strongly dependent on the relative positions of sources and electrodes.

\begin{figure}[!htb]
\begin{center}
\includegraphics[width=10cm]{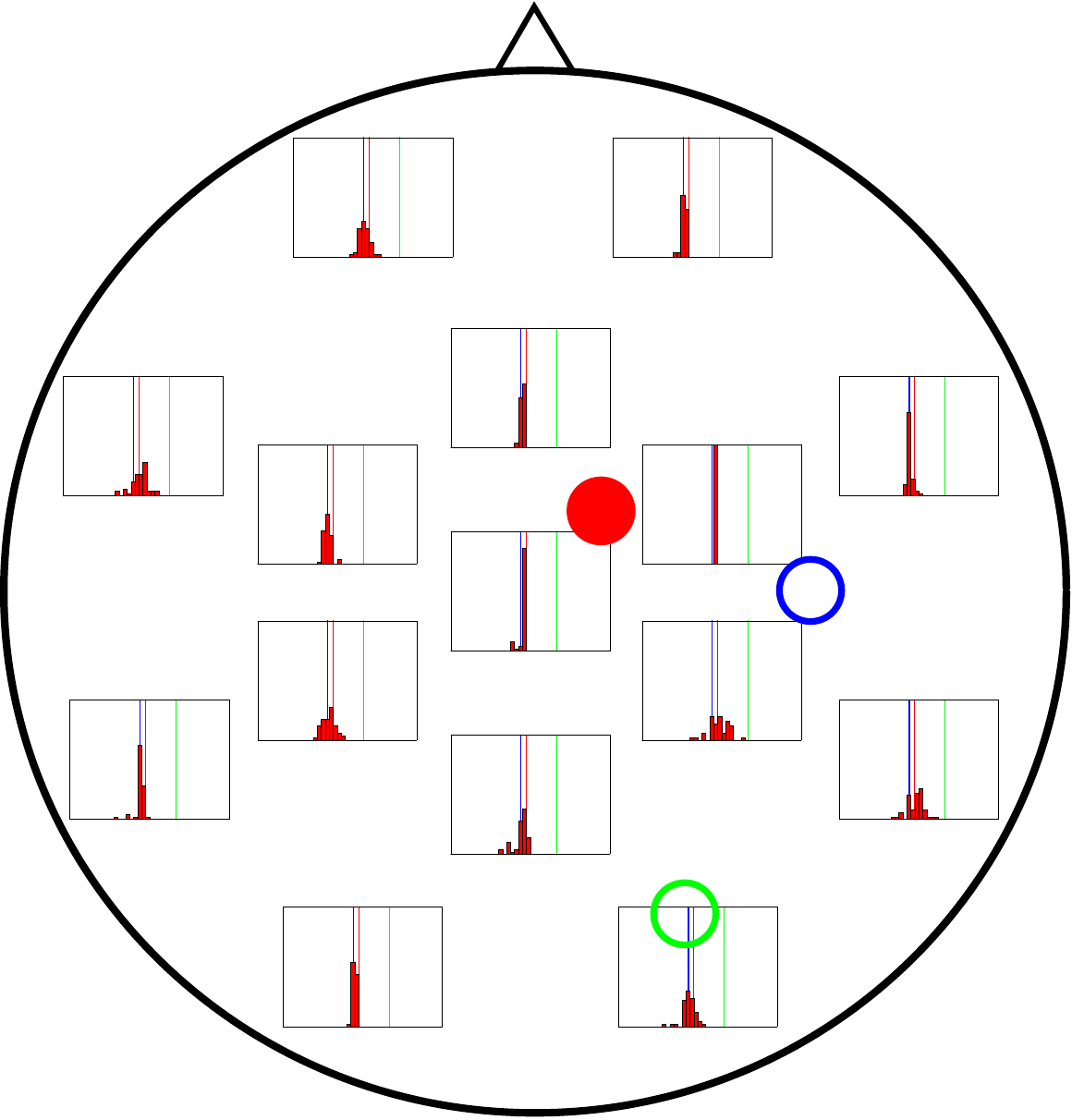}
\caption{{\bf Distribution of 50 realizations of $A$ estimations from a single electrode for cortical column 1 (solid red circle).} The histograms are placed at the location of the corresponding measuring electrode, and the location of the three cortical columns generating the activity are shown with colored circles (with the full circle corresponding to the column whose value of $A$ is being estimated in this figure). Vertical lines with the same colors as the circles mark the corresponding actual $A$ values.
The distributions tend to be narrowest in the vicinities of cortical column 1. Nevertheless, they do not group around the target value of $A_1 = 4.25~mV$ (vertical red line), as they should, but around that of $A_3 = 3.25~mV$ (vertical blue line).}
\label{histograms_1}
\end{center}
\end{figure}

\begin{figure}[!htb]
\begin{center}
\includegraphics[width=10cm]{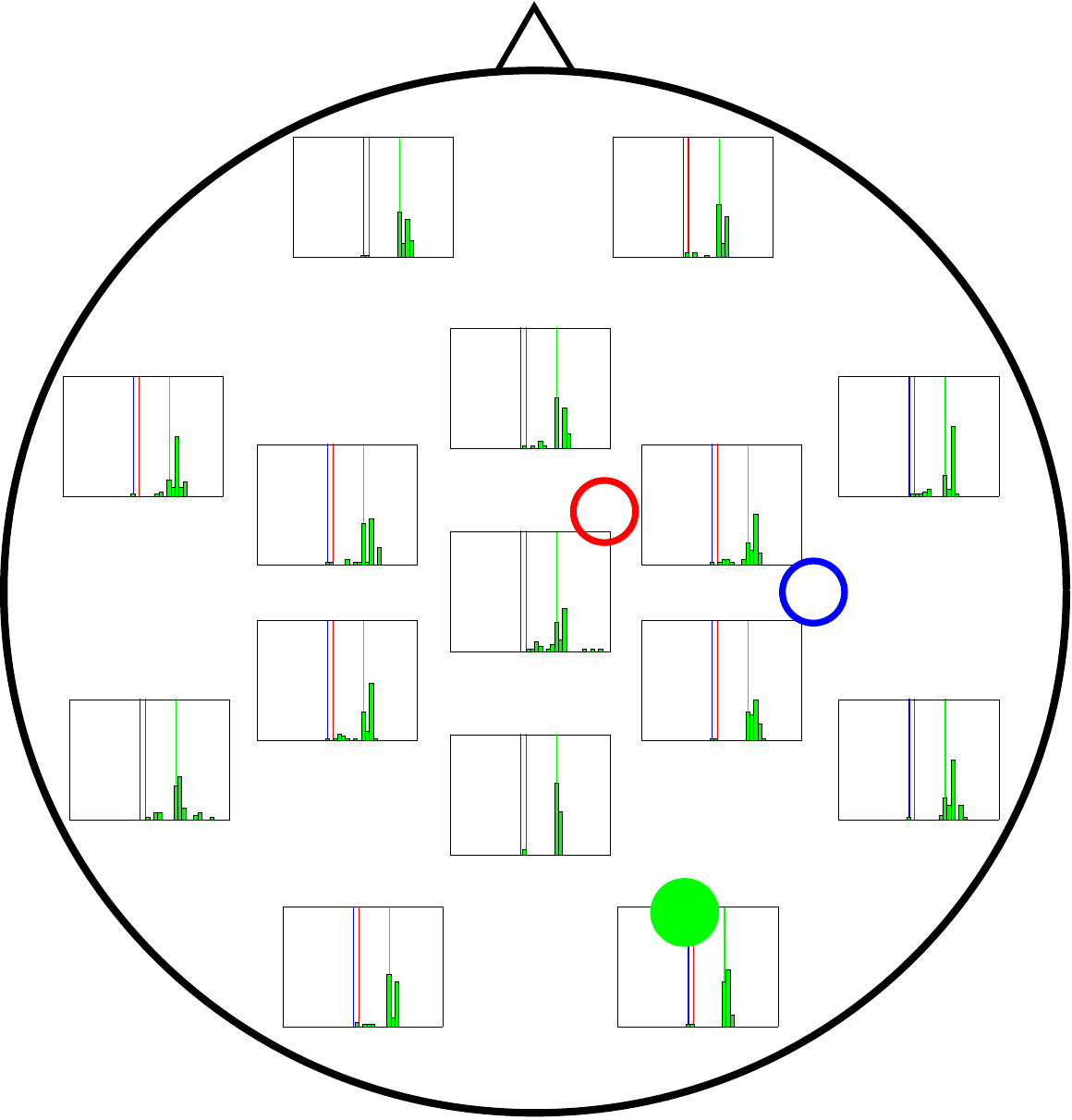}
\caption{{\bf Distribution of 50 realizations of $A$ estimations from a single electrode for cortical column 2 (solid green circle).}
The distributions here are wider than for $A_1$ and $A_3$, although they still tend to be more accurate near the cortical column (solid green circle) and group around the target value of $A_2 = 10.00~mV$ (vertical green line).}
\label{histograms_2}
\end{center}
\end{figure}

\begin{figure}[!htb]
\begin{center}
\includegraphics[width=10cm]{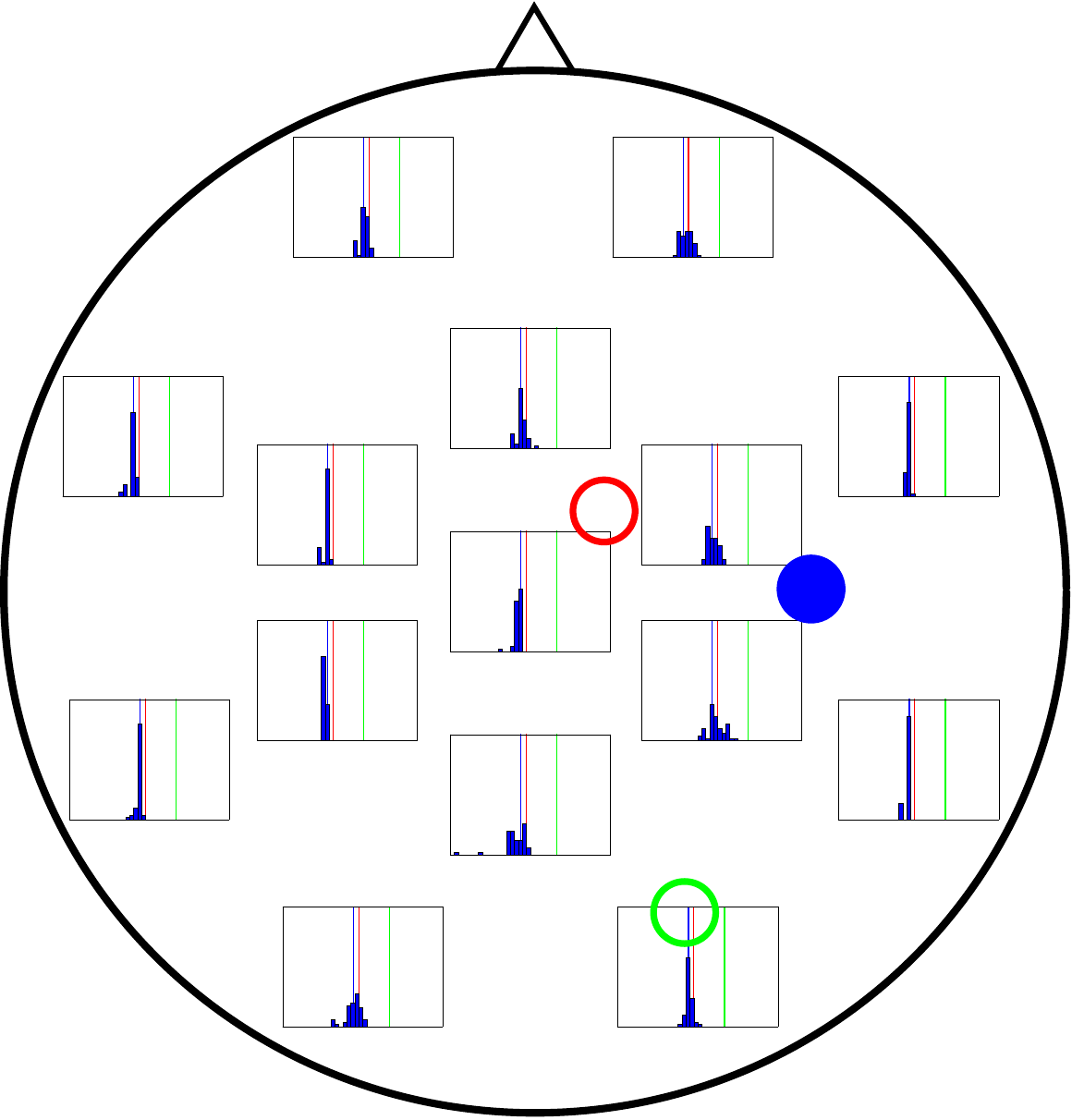}
\caption{{\bf Distribution of 50 realizations of $A$ estimations from a single electrode for cortical column 3 (solid blue circle).}
As in the two previous figures, the distributions for the electrodes closest to the source (solid blue circle) are narrow, grouping around the correct value ($A_3 = 3.25~mV$, vertical blue line). Surprisingly, the electrodes in the far left also give rise to narrow distributions.}
\label{histograms_3}
\end{center}
\end{figure}

In Figs.~\ref{histograms_1} to \ref{histograms_3} we present the results for the estimation of parameter $A$ of each of the three cortical columns separately. The histograms show the distribution of the 50 estimations of $A$ using each electrode, placed in the respective position of the electrode in question. Vertical colored lines in the histograms mark the value of the three $A$ parameters being estimated (one in each figure).
The histograms show a strong dependence on space of the quality of the estimations. As a general trait, the estimations are better when the electrodes are near the cortical column whose value of $A$ is being estimated, whereas the more distant electrodes show a wider distribution of final values for the parameter.

In Fig.~\ref{histograms_1} the distribution of the estimations of $A_1$ are shown. The distributions tend to be narrowest in the vicinities of the cortical column whose $A$ value is being estimated. However, it is noteworthy that the histograms obtained from the observations in distant electrodes tend to group not around the actual value of $A_1 = 4.25~mV$ (red vertical line), but of $A_3 = 3.25~mV$ (blue vertical line). This result suggests that the algorithm is unable to distinguish the origin of the EEG activity when sources and electrodes are distant from each other.

Figure~\ref{histograms_2} shows the results of the estimation of $A_2$ (actual value shown by vertical green lines), revealing wider distributions in general, which indicates a stronger dependence on initial conditions. Although it is true that the electrodes near cortical column 2 perform better in estimating $A$ for that column, the difference with more distant electrodes is not as large as for the estimates of $A$ for cortical columns 1 and 3.

Finally, Fig.~\ref{histograms_3} shows the performance of each electrode when $A_3$ is being estimated (actual value shown by vertical blue lines in the figure). Interestingly, even the electrodes located at the far left of the figure lead to a good estimate of $A$, comparable to that coming from the electrodes in the far right, which are closer to column 3 and could therefore be expected to provide a much more accurate estimation.

However accurate some of the single electrodes' estimations are, using the complete set of 15 electrodes invariably yields better results. This is because, in Kalman filtering, combining many sources of information always improves the final estimation, even if some of the sources are inaccurate or incomplete~\cite{schiff2012neural}.

\subsection*{Three bidirectionally coupled cortical columns: fine parameter estimation}

\begin{figure}[!htb]
\begin{center}
\includegraphics[width=7cm]{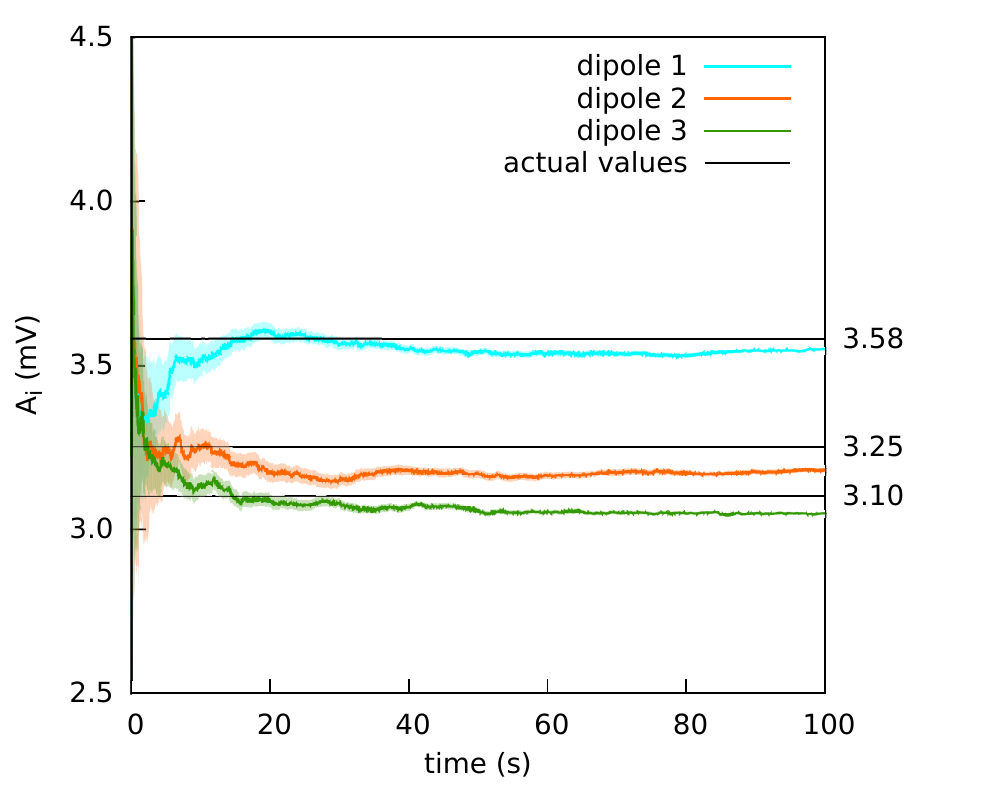}
\caption {{\bf Extracranial fit with parameters close together in value.}
The estimations of the amplitude of the EPSPs of the three cortical columns are shown after averaging over 50 realizations (solid lines); the shadowed areas indicate the standard deviation. The actual values of the amplitudes of the EPSPs are $A_1 = 3.58~mV$, $A_2 = 3.25~mV$, and $A_3 = 3.10~mV$; the rest of the parameters were set to standard values (Table~\ref{table:parameters}). The external input $p$ for each of the three cortical columns is of mean $200~s^{-1}$ and standard deviation $100~s^{-1}$. The coupling constant was set to $k=5$. The noise in the extracranial measurements is of standard deviation $100~mV$. The estimation of the parameters is fairly accurate.}
\label{exp_10}
\end{center}
\end{figure}

In the previous section, the value of $A$ of one of the cortical columns was much larger than the other two.
We now consider the same coupling motif, but with values of the $A$ parameter that are much closer together in value: $A_1 = 3.58~mV$, $A_2 = 3.25~mV$, and $A_3 = 3.10~mV$. The purpose of this test was to ascertain whether the filter could differentiate between parameters with smaller differences in value. This ability is very important if we expect to use the technique in clinical applications. Fig.~\ref{exp_10} shows the extracranial estimation of the $A$ parameters using the complete subset of 15 electrodes. The estimations converge to the actual values with enough accuracy as to give hopes of using the filter in a clinical setting.

\section*{Discussion}

The most important limitation of current data assimilation processes in neuroscience is that the appropriate experimental recordings are usually intracranial.
Using Kalman filtering to fit these data to neural mass models shows promise in several contexts and applications. Here we have modified this type of approach by extending the base neural mass model with a head model, with the aim of integrating non-invasive experimental recordings. We have explored the limitations and advantages of our model using {\it in silico} data in very well controlled conditions. Even though we keep the exploration of the technique using real EEG experimental data in mind, here we bring forth a proof of concept by performing several experiments that address different aspects of the method.

In this paper we have considered a system comprised of three cortical columns, modelled according to Jansen and Rit's equations and coupled following two different motifs. The cortical columns are all driven by a noisy input. The signal from the cortical columns is then transferred to the skull, after which it is corrupted by noise to simulate electrode readings from EEG. These are then used to estimate the amplitude of the excitatory post-synaptic potentials using the Unscented Kalman Filter.

%\subsection*{Three unidirectionally coupled cortical columns}
The first study involves three columns that are coupled unidirectionally with no backflow. The first cortical column is made hyperexcitable by increasing the excitatory post-synaptic potential to $A_1=3.58~mV$; this cortical column causes the second cortical column and, indirectly, the third to modify their behavior as a result of the spiking of the first. For the intracranial estimations, single intracortical electrodes measured the evolution of the three cortical columns independently; for the extracranial estimations, 15 extracranial electrodes were used simultaneously.
Applying the Kalman filter to the extracranial data provided a good estimation of the $A$ parameters; the intracortical measurements, however, yielded mixed results. The estimation for cortical column 1 was accurate, whereas for cortical columns 2 and 3 the estimation was above the target value. We attribute this to the fact that columns 2 and 3 are excited by column 1, which spikes due to a higher value of
$A$. As a consequence, when independently evaluated, the estimation is higher than the actual value. Therefore we suggest that one intracranial electrode provides only a partial view of the system, and thus cannot capture the behaviors of all three cortical columns and the interactions between them; the use of many electrodes provides a more complete view of the system.

%\subsection*{Three bidirectionally coupled cortical columns: coarse parameter estimation}
Next we considered a situation in which the dipoles were coupled bidirectionally in an all-to-all configuration.
The $A$ parameters were chosen such as to cause different dynamic behaviors in the cortical columns: $A_1 = 4.25~mV$, $A_2 = 10.00~mV$, and $A_3 = 3.25~mV$.
Three types of fitting via Kalman filtering were performed, using (i) independent intracortical recordings of single cortical columns were filtered, (ii) the complete subset of 15 extracranial electrodes, and (iii) single extracranial electrodes. The intracortical data were corrupted with two different levels (medium and high) of measurement noise. For both cases, the multi-electrode extracraneal estimation surpasses the intracortical results in both speed and quality; the difference, however, is more marked in the presence of higher measurement noise in the intracortical recordings. The results for the single electrodes show a significant influence of space on the quality of the estimations, in the sense that estimations of electrodes close to the source are relatively
accurate, and electrodes further away from the source might not allow to discriminate the source of the information correctly, or might completely fail to represent the system.

%\subsection*{Three bidirectionally coupled cortical columns: fine parameter estimation}
Finally, we considered the situation of an identical cortical column configuration ---in terms of situation and coupling---, except for the values of the EPSPs of the cortical columns: $A_1 = 3.58~mV$, $A_2 = 3.25~mV$, and $A_3 = 3.10~mV$. This dataset was filtered only extracranially, with the purpose of evaluating the filter's ability to discriminate parameter values within narrower ranges. The results were accurate, which is promising in views of applying the algorithm in a clinical setting.

Taken as a whole, our results show that, independently of the need to explore more realistic situations, extracranial EEG recordings constitute a good candidate to be used together with neural mass models and Kalman filters, provided the method is extended with a head model. Using non-invasive techniques in these processes widens the applications of Kalman-based data assimilation methods in neuroscience.

The adaptation of the method to a specific possible applications deserves its own exploration.
The possibility of determining parameters of cerebral dynamics in a non-invasive manner would allow us to study, for instance, the origins of the variability in EEG recordings. It would also enable exploring automatic biometric-based user recognition systems\cite{Campisi2014} and, through single-patient characterization, tracking the changes in brain dynamics due to aging\cite{Anokhin1996,Yang2017}, and monitoring the evolution of diseases \cite{Soekadar2015}.
The possibility of tracking the evolution of brain states during motor imagery-control \cite{Zich2015} or task-switching control\cite{Phillips2014} is also open. Besides, a good description of the brain state would allow the efficient control of epilepsy\cite{Shan2016}, a good performance in brain-machine interface tasks \cite{DelRMillan2008}, and the detection and control of transcranial brain stimulation \cite{Krause2013}. Rehabilitation tasks \cite{Ara15,Stephan2015} may also benefit from the possibility of monitoring brain states reliably.

\section*{Supporting Information}
\subsection*{S1 Appendix}
\label{s1_appendix}
{\bf The Unscented Kalman Filter (UKF) algorithm.}

UKF is a predictor-corrector algorithm that estimates the state and parameters at a given time step $k$ in two phases. The first one predicts the state based solely on the dynamical information of the system, i.e., the model. The second incorporates a measurement with which to correct the first estimation.

The first step of the algorithm involves computing the expectation of the state and of the state covariance at time instant $k+1$, known as the \textit{a priori} estimation. For this we use a numerical implementation (using Heun's solver) of Jansen and Rit's model of a cortical column~\cite{Jan95, Jan93}, as described in the~\nameref{methods} section.

The nature of the nonlinearities of this model prevents us from using a simple linearization approach to propagating the statistics of the state variables across the transformation. Therefore, we incorporate the unscented transform (UT) in our formulation of the Kalman filter, which, instead of attempting to propagate a distribution through the nonlinearity, first propagates a series of deterministically chosen points through the nonlinearity and then recovers the statistical information of the distribution from these.

Therefore, the \textit{a priori} estimation of the state, $\bm{\hat{x}_k^-}$, is obtained as follows, beginning with the calculation and projection of the $2n+1$ (where $n$ is the state size) sigma points,
\begin{align}\label{eq:sigma_points}
\begin{split}
\bm{\Sigma_{k-1,0}} &= \bm{\hat{x}_{k-1}} \\
\bm{\Sigma_{k-1,i}} &= \bm{\hat{x}_{k-1}} + \left(\sqrt{(n + \lambda)\bm{P_{k-1}}}\right)_i,\quad i=1,...,n \\
\bm{\Sigma_{k-1,i}} &= \bm{\hat{x}_{k-1}} - \left(\sqrt{(n + \lambda)\bm{P_{k-1}}}\right)_{i-n},\quad i=n+1,...,2n
\end{split}
\end{align}
where $\bm{P_{k-1}}$ is the estimated state covariance matrix for the previous time step. This continues with the condensation of the projected sigma points into the \textit{a priori} state estimate:
\begin{align}\label{eq:a_priori_x}
\bm{X_{k|k-1}^*} &= f(\bm{\Sigma_{k-1}}, \bm{u_{k-1}}) \\
\bm{\hat{x}_k^-} &= \sum_{i = 0}^{2L}W_i^m\ \bm{X_{i, k|k-1}^*}\\
\bm{P_k^-} &= \sum_{i = 0}^{2L} W_i^{cov} [\bm{X_{i, k|k-1}^*} - \bm{\hat{x}_k^-}] [\bm{X_{i, k|k-1}^*} - \bm{\hat{x}_k^-}]^T + \bm{Q}
\end{align}
where $\bm{Q}$ is the state error covariance and $\bm{W^m}$ and $\bm{W^{cov}}$ are the weight vectors, defined as
\begin{align}\label{eq:weights}
\begin{split}
W_0^m &= \frac{\lambda}{n+\lambda} \\
W_0^{cov} &= \frac{\lambda}{n+\lambda} + 1 - \alpha^2 + \beta \\
W_i^m = W_i^{cov} &= \frac{1}{2(n+\lambda)}, i = 1,...,2n
\end{split}
\end{align}

In Eqs.~\ref{eq:sigma_points} and~\ref{eq:weights}, $\alpha$, $\beta$ and $\kappa$ are scaling factors, and $\lambda$ is calculated as $\lambda = \alpha^2(n+\kappa)-n$. $\alpha$, the primary scaling factor, determines the spread of the sigma points around the mean and is usually set between 0.001 and 1. $\beta$ contains prior information about the distribution of $\bm{x}$; for Gaussian distributions, its optimal value is 2. $\kappa$, the tertiary scaling parameter, is usually set to 0~\cite{Mer00}.

Finally, the sigma points are redrawn~\cite{Mer01} and the estimation of the measurement, $\bm{\hat{y}_k^-}$, is calculated:
\begin{align}\label{eq:measurement_estimation}
\bm{\Upsilon_{k|k-1}} &= \bm{H}[\bm{\Sigma_{k|k-1}}] \\
\bm{\hat{y}_k^-} &= \sum_{i = 0}^{2L} W_i^m\ \bm{\Upsilon_{i, k|k-1}}
\end{align}

The use of a measurement to correct the state estimation implies the mapping of the \textit{a priori} estimate onto the measurement space for comparison. It is worthwhile to note that, in our case, this transformation is a linear matrix $\bm{H}$ which relates the state of the cortical columns to an EEG reading. See~\nameref{head} section for details.

The second step of the algorithm corrects the \textit{a priori} estimation of state and covariance by using the information available from the most recent measurement (in our case, an EEG reading). The impact of the measurement is determined by the Kalman gain $\bm{K_k}$, which essentially expresses the level of confidence on the accuracy of the model and the level of noise in the data.

\begin{align}\label{eq:a_posteriori_x}
\bm{P_{y_k y_k}} &= \sum_{i = 0}^{2L} W_i^{cov}\ [\bm{\Upsilon_{i, k|k-1}} - \bm{\hat{y}_k^-}] [\bm{\Upsilon_{i, k|k-1}} - \bm{\hat{y}_k^-}]^T + \bm{R} \\
\bm{P_{x_k y_k}} &= \sum_{i = 0}^{2L} W_i^{cov}\ [\bm{X_{i, k|k-1}} - \bm{\hat{x}_k^-}] [\bm{\Upsilon_{i, k|k-1}} - \bm{\hat{y}_k^-}]^T \\
\bm{K_k} &= \bm{P_{x_k y_k}}\ \bm{P_{y_k y_k}}^{-1}\\
\bm{\hat{x}_k} &= \bm{\hat{x}_k^-} + \bm{K_k}(\bm{z_k} - \bm{\hat{y}_k^-}) \\
\bm{P_k} &= \bm{P_k^-} - \bm{K_k}\ \bm{P_{y_k y_k}}\ \bm{K_k}^T
\end{align}
where $\bm{P_{y_k y_k}}$ is the predicted measurement covariance, $\bm{P_{x_k y_k}}$ is the state-measurement cross-covariance, $\bm{R}$ is the measurement error covariance, and $\bm{z_k}$ is the measurement for the current time step.

\section*{Acknowledgments}

This work was partially supported by the Catalan Government (AGAUR grant FI-DGR 2014-2017) and by the Spanish Ministry of Economy and Competitiveness and FEDER (project {FIS2015-66503}). JGO also acknowledges support from the the Catalan Government (project 2014SGR0947), the ICREA Academia programme, and from the ``Mar\'ia de Maeztu'' programme for Units of Excellence in R\&D (Spanish Ministry of Economy and Competitiveness, MDM-2014-0370).

\nolinenumbers

\bibliography{kalman}

\begin{thebibliography}{10}

\bibitem{Yuste2015}
Yuste R.
\newblock {From the neuron doctrine to neural networks}.
\newblock Nature Reviews Neuroscience. 2015;16(8):487--497.
\newblock Available from:
  \url{http://www.nature.com/doifinder/10.1038/nrn3962}.

\bibitem{Braitenberg91a}
Braitenberg V, Sch{\"u}tz A.
\newblock Anatomy of the Cortex.
\newblock No.~18 in Studies of Brain Function. Springer; 1991.
\newblock Statistics and Geometry.

\bibitem{Izhikevich2008}
Izhikevich EM, Edelman GM.
\newblock {Large-scale model of mammalian thalamocortical systems}.
\newblock Proceedings of the National Academy of Sciences.
  2008;105(9):3593--3598.
\newblock Available from:
  \url{http://www.pnas.org/cgi/doi/10.1073/pnas.0712231105}.

\bibitem{Callaway2016}
Callaway EM.
\newblock {Micro-, Meso- and Macro-Connectomics of the Brain}.
\newblock Springer; 2016.
\newblock Available from:
  \url{http://link.springer.com/10.1007/978-3-319-27777-6}.

\bibitem{Fre87}
Freeman WJ.
\newblock {Simulation of chaotic EEG patters with a dynamic model of the
  olfactory system}.
\newblock {Biological Cybernetics}. 1987;56:139--150.

\bibitem{Faure2001}
Faure P, Korn H.
\newblock {Is there chaos in the brain? I. Concepts of nonlinear dynamics and
  methods of investigation.}
\newblock C R Acad Sci III. 2001 sep;324(9):773--93.
\newblock Available from: \url{http://www.ncbi.nlm.nih.gov/pubmed/11558325}.

\bibitem{Shadlen1994}
Shadlen MN, Newsome WT.
\newblock {Noise, neural codes and cortical organization}.
\newblock Current Opinion in Neurobiology. 1994;4(4):569--579.

\bibitem{Faisal2008}
Faisal AA, Selen LPJ, Wolpert DM.
\newblock {Noise in the nervous system}.
\newblock Nature Reviews Neuroscience. 2008;9(4):292--303.
\newblock Available from:
  \url{http://www.nature.com/doifinder/10.1038/nrn2258}.

\bibitem{Schiff1994}
Schiff SJ, Jerger K, Duong DH, Chang T, Spano ML, Ditto WL.
\newblock {Controlling chaos in the brain.}
\newblock Nature. 1994;370(6491):615--620.

\bibitem{VanVreeswijk1996}
van Vreeswijk C, Sompolinsky H.
\newblock {Chaos in neuronal networks with balanced excitatory and inhibitory
  activity.}
\newblock Science (New York, NY). 1996;274(5293):1724--6.
\newblock Available from: \url{http://www.ncbi.nlm.nih.gov/pubmed/8939866}.

\bibitem{Celletti1999}
Celletti A, Froeschle C, Tetko IV, Villa AEP.
\newblock {Deterministic behaviour of short time series}.
\newblock Meccanica. 1999;34(3):147--154.
\newblock Available from:
  \url{http://links.isiglobalnet2.com/gateway/Gateway.cgi?GWVersion=2{\&}SrcAuth=mekentosj{\&}SrcApp=Papers{\&}DestLinkType=FullRecord{\&}DestApp=WOS{\&}KeyUT=000083554700001{\%}5Cnpapers3://publication/uuid/9EE31316-A232-4916-8227-D1E9D35B4355}.

\bibitem{Stam2005}
Stam CJ.
\newblock {Nonlinear dynamical analysis of EEG and MEG: review of an emerging
  field.}
\newblock Clin Neurophysiol. 2005 oct;116(10):2266--301.
\newblock Available from: \url{http://www.ncbi.nlm.nih.gov/pubmed/16115797}.

\bibitem{Buzsaki2012}
Buzs{\'{a}}ki G, Anastassiou Ca, Koch C.
\newblock {The origin of extracellular fields and currents--EEG, ECoG, LFP and
  spikes.}
\newblock Nature reviews Neuroscience. 2012;13(6):407--20.
\newblock Available from: \url{http://www.ncbi.nlm.nih.gov/pubmed/22595786}.

\bibitem{Wright1996}
Wright JJ, Liley DTJ.
\newblock Dynamics of the Brain at Global and Microscopic Scales: Neural
  Networks and the EEG.
\newblock Behavioral and Brain Sciences. 1996;19(2):285.

\bibitem{Rabinovich2006}
Rabinovich MI, Varona P, Selverston AI, Abarbanel HDI.
\newblock {Dynamical principles in neuroscience}.
\newblock Reviews of Modern Physics. 2006;78(4).

\bibitem{Eliasmith2012}
Eliasmith C, Stewart TC, Choo X, Bekolay T, DeWolf T, Tang C, et~al.
\newblock {A Large-Scale Model of the Functioning Brain}.
\newblock Science. 2012;338(6111):1202--1205.
\newblock Available from:
  \url{http://www.sciencemag.org/cgi/doi/10.1126/science.1225266{\%}5Cnhttp://www.ncbi.nlm.nih.gov/pubmed/23197532}.

\bibitem{Deco2014}
Deco G, Ponce-Alvarez A, Hagmann P, Romani G, Mantini D, Corbetta M.
\newblock {How local excitation-inhibition ratio impacts the whole brain
  dynamics}.
\newblock The Journal of neuroscience. 2014;34(23):7886--98.
\newblock Available from:
  \url{http://www.pubmedcentral.nih.gov/articlerender.fcgi?artid=4044249{\&}tool=pmcentrez{\&}rendertype=abstract}.

\bibitem{David2003b}
David O, Friston KJ.
\newblock {A neural mass model for MEG/EEG: Coupling and neuronal dynamics}.
\newblock NeuroImage. 2003;20(3):1743--1755.

\bibitem{Grimbert2006a}
Grimbert F, Faugeras O.
\newblock {Bifurcation analysis of Jansen's neural mass model.}
\newblock Neural computation. 2006;18(12):3052--3068.

\bibitem{Cona2011a}
Cona F, Zavaglia M, Massimini M, Rosanova M, Ursino M.
\newblock {A neural mass model of interconnected regions simulates rhythm
  propagation observed via TMS-EEG}.
\newblock NeuroImage. 2011;57(3):1045--1058.

\bibitem{Coombes2010}
Coombes S.
\newblock {Large-scale neural dynamics: Simple and complex}.
\newblock NeuroImage. 2010;52(3):731--739.

\bibitem{Babajani2006a}
Babajani A, Soltanian-Zadeh H.
\newblock {Integrated MEG/EEG and fMRI model based on neural masses}.
\newblock IEEE Transactions on Biomedical Engineering. 2006;53(9):1794--1801.

\bibitem{Babiloni2005}
Babiloni F, Cincotti F, Babiloni C, Carducci F, Mattia D, Astolfi L, et~al.
\newblock {Estimation of the cortical functional connectivity with the
  multimodal integration of high-resolution EEG and fMRI data by directed
  transfer function}.
\newblock NeuroImage. 2005;24(1):118--131.

\bibitem{Bojak2010}
Bojak I, Oostendorp TF, Reid AT, K{\"{o}}tter R.
\newblock {Connecting mean field models of neural activity to EEG and fMRI
  data}.
\newblock Brain Topography. 2010;23(2):139--149.

\bibitem{Jan95}
Jansen BH, Rit VG.
\newblock {Electroencephalogram and visual evoked potential generation in a
  mathematical model of coupled cortical columns}.
\newblock {Biological Cybernetics}. 1995;73(4):357--366.

\bibitem{David2005}
David O, Harrison L, Friston KJ.
\newblock {Modelling event-related responses in the brain}.
\newblock NeuroImage. 2005;25(3):756--770.

\bibitem{Spiegler2011}
Spiegler A, Kn{\"{o}}sche TR, Schwab K, Haueisen J, Atay FM.
\newblock {Modeling brain resonance phenomena using a neural mass model}.
\newblock PLoS Computational Biology. 2011;7(12).

\bibitem{Wright2001}
Wright JJ, Robinson PA, Rennie CJ, Gordon E, Bourke PD, Chapman CL, et~al.
\newblock {Toward an integrated continuum model of cerebral dynamics: The
  cerebral rhythms, synchronous oscillation and cortical stability}.
\newblock BioSystems. 2001;63(1-3):71--88.

\bibitem{Schiff2008}
Schiff SJ, Sauer T.
\newblock {Kalman filter control of a model of spatiotemporal cortical
  dynamics.}
\newblock Journal of neural engineering. 2008;5(1):1--8.

\bibitem{Kiebel2008}
Kiebel SJ, Garrido MI, Moran RJ, Friston KJ.
\newblock {Dynamic causal modelling for EEG and MEG}.
\newblock Cognitive Neurodynamics. 2008;2(2):121--136.

\bibitem{Shine2015}
Shine JM, Koyejo O, Bell PT, Gorgolewski KJ, Gilat M, Poldrack RA.
\newblock {Estimation of dynamic functional connectivity using Multiplication
  of Temporal Derivatives}.
\newblock NeuroImage. 2015;122:399--407.

\bibitem{Ma2016}
Ma X, Chou CA, Sayama H, Chaovalitwongse WA.
\newblock {Brain response pattern identification of fMRI data using a particle
  swarm optimization-based approach}.
\newblock Brain Informatics. 2016;3(3):181--192.
\newblock Available from:
  \url{http://link.springer.com/10.1007/s40708-016-0049-z}.

\bibitem{Hamilton2014}
Hamilton F, Cressman J, Peixoto N, Sauer T.
\newblock {Reconstructing neural dynamics using data assimilation with multiple
  models}.
\newblock EPL (Europhysics Letters). 2014;107(6):68005.
\newblock Available from:
  \url{http://www.scopus.com/inward/record.url?eid=2-s2.0-84907292009{\&}partnerID=tZOtx3y1}.

\bibitem{Freestone2013b}
Freestone DR, Kuhlmann L, Chong MS, Nesic D, Grayden DB, Aram P, et~al.
\newblock {Patient-specific neural mass modeling - stochastic and deterministic
  methods}.
\newblock Recent Advances in Predicting and Preventing Epileptic Seizures.
  2013;p. 63--82.
\newblock Available from:
  \url{https://hal.archives-ouvertes.fr/hal-00876475/{\%}5Cnhttp://www.worldscientific.com/doi/abs/10.1142/9789814525350{\_}0005}.

\bibitem{Freestone2014}
Freestone DR, Karoly PJ, Ne{\v{s}}i{\'{c}} D, Aram P, Cook MJ, Grayden DB.
\newblock {Estimation of effective connectivity via data-driven neural
  modeling.}
\newblock Frontiers in neuroscience. 2014;8(November):383.
\newblock Available from:
  \url{http://www.pubmedcentral.nih.gov/articlerender.fcgi?artid=4246673{\&}tool=pmcentrez{\&}rendertype=abstract}.

\bibitem{Cao2015}
Cao Y, Ren K, Su F, Deng B, Wei X, Wang J.
\newblock {Suppression of seizures based on the multi-coupled neural mass
  model}.
\newblock Chaos. 2015;25(10).

\bibitem{Bojak2005}
Bojak I, Liley D.
\newblock Modeling the effects of anesthesia on the electroencephalogram.
\newblock Physical Review E. 2005;71(4):041902.

\bibitem{Kuhlmann2016}
Kuhlmann L, Freestone DR, Manton JH, Heyse B, Vereecke HEM, Lipping T, et~al.
\newblock {Neural mass model-based tracking of anesthetic brain states}.
\newblock NeuroImage. 2016;133:438--456.

\bibitem{Zhang1995}
Zhang Z.
\newblock {A fast method to compute surface potentials generated by dipoles
  within multilayer anisotropic spheres.}
\newblock Physics in medicine and biology. 1995;40(3):335--349.

\bibitem{Mosher1999}
Mosher JC, Leahy RM, Lewis PS.
\newblock {EEG and MEG: forward solutions for inverse methods}.
\newblock IEEE Transactions on Biomedical Engineering. 1999;46(3):245--259.

\bibitem{Haufe2011}
Haufe S, Tomioka R, Dickhaus T, Sannelli C, Blankertz B, Nolte G, et~al.
\newblock {Large-scale EEG/MEG source localization with spatial flexibility}.
\newblock NeuroImage. 2011;54(2):851--859.

\bibitem{Verhellen2007}
Verhellen E, Boon P.
\newblock {EEG source localization of the epileptogenic focus in patients with
  refractory temporal lobe epilepsy, dipole modelling revisited.}
\newblock Acta neurologica Belgica. 2007;107(3):71--77.

\bibitem{Gotman2003}
Gotman J.
\newblock {Noninvasive methods for evaluating the localization and propagation
  of epileptic activity.}
\newblock Epilepsia. 2003;44 Suppl 1:21--29.

\bibitem{Jan93}
Jansen BH, Zouridakis G, Brandt ME.
\newblock {A neurophysiologically-based mathematical model of flash visual
  evoked potentials}.
\newblock {Biological Cybernetics}. 1993;68:275--283.

\bibitem{Mer01}
Merwe RVD, Wan EA.
\newblock {The square-root unscented Kalman filter for state and
  parameter-estimation}.
\newblock In: 2001 IEEE International Conference on Acoustics, Speech, and
  Signal Processing. Proceedings. (ICASSP '01). vol.~6; 2001. p. 3461--3464.

\bibitem{Jul04}
Julier SJ, Uhlmann JK.
\newblock {Unscented Filtering and Nonlinear Estimation}.
\newblock {Proceedings of the IEEE}. 2004;92(3):401--422.

\bibitem{Kal60}
{Rudolf Emil Kalman}.
\newblock {A New Approach to Linear Filtering and Prediction Problems}.
\newblock {Transactions of the ASME--Journal of Basic Engineering}.
  1960;82(Series D):35--45.

\bibitem{Lop74}
Silva FHLD, Hoeks A, Smits H, Zetterberg LH.
\newblock {Model of brain rhythmic activity}.
\newblock {Kybernetic}. 1974;15(1):27--37.

\bibitem{Fau09}
Faugeras O, Touboul J, Cessac B.
\newblock {A constructive mean-field analysis of multi-population neural
  networks with random synaptic weights and stochastic inputs}.
\newblock {Frontiers in Computational Neuroscience}. 2009;3(1).

\bibitem{Lop10}
Silva FLD.
\newblock 5.
\newblock In: {EEG: Origin and Measurement}. {World Scientific Publishing Co.};
  2011. p. 63--82.
\newblock Available from:
  \url{http://www.worldscientific.com/doi/abs/10.1142/9789814525350_0005}.

\bibitem{Pons2010}
Pons AJ, Cantero JL, Atienza M, Garcia-Ojalvo J.
\newblock {Relating structural and functional anomalous connectivity in the
  aging brain via neural mass modeling}.
\newblock NeuroImage. 2010;52(3):848--861.

\bibitem{Ary81}
Ary JP, Klein SA, Fender DH.
\newblock {Location of Sources of Evoked Scalp Potentials: Corrections for
  Skull and Scalp Thicknesses}.
\newblock {IEEE Transactions on Biomedical Engineering}.
  1981;BME-28(6):447--452.

\bibitem{Jurcak2007}
Jurcak V, Tsuzuki D, Dan I.
\newblock {10/20, 10/10, and 10/5 systems revisited: Their validity as relative
  head-surface-based positioning systems}.
\newblock NeuroImage. 2007;34(4):1600--1611.

\bibitem{Ber94}
Berg P, Scherg M.
\newblock {A fast method for forward computation of multiple-shell spherical
  head models}.
\newblock {Electroencephalography and Clinical Neurophysiology}.
  1994;90:58--64.

\bibitem{Mer00}
Merwe RVD, Wan EA.
\newblock {The Unscented Kalman Filter for Nonlinear Estimation}.
\newblock In: Adaptive Systems for Signal Processing, Communications, and
  Control Symposium 2000. AS-SPCC. The IEEE 2000; 2000. p. 153--158.

\bibitem{Solonen2014}
Solonen A, Hakkarainen J, Ilin A, Abbas M, Bibov A.
\newblock {Estimating model error covariance matrix parameters in extended
  Kalman filtering}.
\newblock Nonlinear Processes in Geophysics. 2014;21(5):919--927.

\bibitem{Ric04}
Richards JE.
\newblock {Recovering dipole sources from scalp-recorded
  event-related-potentials using component analysis: principal component
  analysis and independent component analysis}.
\newblock {International Journal of Psychophysiology}. 2004;54:201--220.

\bibitem{Toral2014}
Toral R, Colet P.
\newblock Stochastic numerical methods: an introduction for students and
  scientists.
\newblock John Wiley \& Sons; 2014.

\bibitem{Liu2013}
Liu X, Gao Q.
\newblock Parameter estimation and control for a neural mass model based on the
  unscented Kalman filter.
\newblock Phys Rev E. 2013 Oct;88:042905.
\newblock Available from:
  \url{http://link.aps.org/doi/10.1103/PhysRevE.88.042905}.

\bibitem{Gri06}
Grimbert F, Faugeras O.
\newblock {Bifurcation analysis of Jansen's neural mass model}.
\newblock {Neural Computation}. 2006;18(12):3052--3068.

\bibitem{Fre14}
Freestone DR, Karoly PJ, Ne{\v s}i{\'c} D, Aram P, Cook MJ, Grayden DB.
\newblock Estimation of effective connectivity via data-driven neural modeling.
\newblock Frontiers in Neuroscience. 2014;8:383.
\newblock Available from:
  \url{http://journal.frontiersin.org/article/10.3389/fnins.2014.00383}.

\bibitem{schiff2012neural}
Schiff SJ.
\newblock Neural Control Engineering: The Emerging Intersection Between Control
  Theory and Neuroscience.
\newblock Computational neuroscience. MIT Press; 2012.
\newblock Available from: \url{https://books.google.es/books?id=P9UvTQtnqKwC}.

\bibitem{Campisi2014}
Campisi P, Rocca DL.
\newblock {Brain waves for automatic biometric-based user recognition}.
\newblock IEEE Transactions on Information Forensics and Security.
  2014;9(5):782--800.

\bibitem{Anokhin1996}
Anokhin AP, Birbaumer N, Lutzenberger W, Nikolaev A, Vogel F.
\newblock {Age increases brain complexity}.
\newblock Electroencephalography and Clinical Neurophysiology.
  1996;99(1):63--68.

\bibitem{Yang2017}
Yang S, Deng B, Wang J, Li H, Liu C, Fietkiewicz C, et~al.
\newblock {Efficient implementation of a real-time estimation system for
  thalamocortical hidden Parkinsonian properties}.
\newblock Scientific Reports. 2017;.

\bibitem{Soekadar2015}
Soekadar SR, Birbaumer N, Slutzky MW, Cohen LG.
\newblock {Brain-machine interfaces in neurorehabilitation of stroke}.
\newblock Neurobiology of Disease. 2015;83:172--179.

\bibitem{Zich2015}
Zich C, Debener S, Kranczioch C, Bleichner MG, Gutberlet I, {De Vos} M.
\newblock {Real-time EEG feedback during simultaneous EEG-fMRI identifies the
  cortical signature of motor imagery}.
\newblock NeuroImage. 2015;114:438--447.

\bibitem{Phillips2014}
Phillips JM, Vinck M, Everling S, Womelsdorf T.
\newblock {A long-range fronto-parietal 5- to 10-Hz network predicts "top-down"
  controlled guidance in a task-switch paradigm}.
\newblock Cerebral Cortex. 2014;24(8):1996--2008.

\bibitem{Shan2016}
Shan B, Wang J, Deng B, Wei X, Yu H, Zhang Z, et~al.
\newblock {Particle swarm optimization algorithm based parameters estimation
  and control of epileptiform spikes in a neural mass model}.
\newblock Chaos. 2016;26.

\bibitem{DelRMillan2008}
{Del R  Mill{\'{a}}n} J, Ferrez PW, Gal{\'{a}}n F, Lew E, Chavarriaga R.
\newblock {Non-Invasive Brain-Machine Interaction}.
\newblock International Journal of Pattern Recognition and Artificial
  Intelligence. 2008;22(05):959--972.

\bibitem{Krause2013}
Krause B, M{\'{a}}rquez-Ruiz J, {Cohen Kadosh} R.
\newblock {The effect of transcranial direct current stimulation: a role for
  cortical excitation/inhibition balance?}
\newblock Frontiers in human neuroscience. 2013;7(September):602.
\newblock Available from:
  \url{http://www.pubmedcentral.nih.gov/articlerender.fcgi?artid=3781319{\&}tool=pmcentrez{\&}rendertype=abstract}.

\bibitem{Ara15}
Aram P, Freestone DR, Cook MJ, Kadirkamanathan V, Grayden DB.
\newblock Model-based estimation of intra-cortical connectivity using
  electrophysiological data.
\newblock NeuroImage. 2015;118:563 -- 575.
\newblock Available from:
  \url{http://www.sciencedirect.com/science/article/pii/S1053811915005534}.

\bibitem{Stephan2015}
Stephan KE, Schlagenhauf F, Huys QJM, Raman S, Aponte EA, Brodersen KH, et~al.
\newblock {Computational neuroimaging strategies for single patient
  predictions}.
\newblock NeuroImage. 2015;145:180--199.

\end{thebibliography}

\end{document}